\documentclass[aps, prd, article]{revtex4}
\bibliography{apsrev}
\usepackage{graphicx}
\usepackage{pstricks}
\begin{document}

\title{Noncommutative geometry inspired  Schwarzschild black hole}

\author{Piero Nicolini} 
\email{nicolini@cmfd.univ.trieste.it}
  \affiliation{Dipartimento di Matematica e Informatica,
  Universit\`a degli Studi di Trieste, 
 Dipartimento di Matematica, Politecnico di Torino, Turin,
 Istituto Nazionale di Fisica Nucleare, Sezione di Trieste,
 Institut Jo\v{z}ef Stefan, Ljubljana}

        \author{Anais Smailagic}
 \email{anais@ictp.trieste.it} 
 \affiliation{Istituto Nazionale di Fisica Nucleare, Sezione di Trieste}

 \author{Euro Spallucci} 
  \email{spallucci@trieste.infn.it}
   \affiliation{Dipartimento di Fisica Teorica, Universit\`a degli Studi di 
Trieste, and Istituto Nazionale di Fisica Nucleare, Sezione di Trieste}

\begin{abstract}
We investigate the behavior of a noncommutative radiating Schwarzschild black 
hole. It is shown that coordinate
noncommutativity cures usual problems encountered in the description 
of the terminal phase of black hole evaporation. More in detail, we find that:\\
 the evaporation end-point is a zero temperature extremal black hole even
in the case of electrically neutral, non-rotating, objects;\\ 
 there exists a finite \textit{maximum} temperature that the black hole can 
reach before cooling down to \textit{absolute zero};\\ 
 there is no curvature singularity at the origin, rather we obtain a regular
\textit{DeSitter} core at short distance. 
 \end{abstract}
\pacs{04.70.Dy, 02.40.Gh}
\maketitle
 The theoretical discovery of radiating black holes 
\cite{hawking} disclosed the first physically relevant window on
the mysteries of quantum gravity. After thirty years of intensive
research in this field (~see \cite{paddy} for a recent review with
an extensive reference list~) various aspects of the problem still
remain under debate. For instance, a fully satisfactory
description of the late stage of black hole evaporation is still
missing. The string/black hole correspondence principle
\cite{corr} suggests that in this extreme regime stringy effects
cannot be neglected. This is just one of many examples of how
the development of string theory has affected various aspects of
theoretical physics. Among different  outcomes of string theory, we focus
on  the result that target spacetime coordinates become \textit{noncommuting}
operators on a $D$-brane \cite{sw}. Thus, string-brane coupling has put in 
evidence the necessity of \textit{spacetime quantization}. This indication gave 
a new boost to reconsider older ideas of similar kind  pioneered
in a, largely ignored, paper by Snyder \cite{sny}.\\
The noncommutativity of spacetime can be encoded in the commutator

\begin{equation}
\left[\, \mathbf{x}^\mu\ , \mathbf{x}^\nu\, \right]= i \,
\theta^{\mu\nu} \label{ncx}
\end{equation}

where  $\theta^{\mu\nu}$ is an anti-symmetric  matrix which
determines the fundamental cell discretization of spacetime much
in the same way as the Planck constant $\hbar$ discretizes the
phase space.\\
The modifications of  quantum field theory implied by (\ref{ncx})
has been  recently investigated to a large extent. Currently the approach
to noncommutative quantum field theory follows two distinct
paths: one is based on the Weyl-Wigner-Moyal $\ast$-product and
the other on coordinate coherent state formalism \cite{ae0}, \cite{ae2}.
In a recent paper, following the second formulation,  it has been
shown that controversial questions of  Lorentz invariance and
unitary \cite{casino}, raised in the $\ast$-product approach, can
be solved assuming $\theta^{\mu\nu}= \theta\,
\mathrm{diag}\left(\, \epsilon_{ij}\ , \epsilon_{ij} \dots
\,\right)$\cite{ae}, where
        $\theta$ is a constant  with dimension of length squared.
Furthermore,  the coordinate coherent state approach profoundly
modifies the structure of the Feynman propagator rendering the
theory ultraviolet finite, i.e. curing the short distance behavior of pointlike
structures. It is thus reasonable to believe that noncommutativity
could as well cure divergences  that appear, under various forms,
in General Relativity.  In such a framework, it would be  of particular
interest to study the effects of noncommutativity on the terminal phase of 
black hole evaporation.  In the standard formulation, the temperature 
diverges  and an arbitrarily large curvature state is reached.
A preliminary analysis both in $(1+1)$ dimensions \cite{berna} and in $(3+1)$ 
linearized General Relativity \cite{ag}, \cite{piero2005} suggests
that  noncommutativity can cure this pathological short distance  
behavior. \\
In order to prove the above conjecture, and in view of spherical symmetry, it
is sufficient to consider a $r$-$t$ section  of the noncommutative 
Schwarzschild metric. At first glance,one could think of modifying the $4D$ 
Einstein action to incorporate noncommutative effects.  To analyze
black hole evaporation in this new framework one has to solve corresponding 
field equations. 
Fortunately, we shall argue that it is not necessary to change the Einstein tensor
part of the field equations, and that the noncommutative effects can be implemented
acting only on the matter source. The line of reasoning is the following. 
Metric field is a geometrical structure defined over an underlying manifold.
Curvature measures the strength of the metric field, i.e. is the response to the
presence of a mass-energy distribution. What we know for sure, is that
noncommutativity is an \textit{intrinsic property} of the manifold itself, rather
than a super-imposed geometrical structure. Under this respect, it affects
gravity in a subtle, indirect way. By studying quantum mechanics and quantum
field theory in simple non-commutative manifolds, e.g. a plane,
one sees that noncommutativity influences matter energy and momentum distribution
and propagation \cite{ae0}, \cite{ae2}, \cite{noi4} . On the other hand, 
energy-momentum density determines spacetime curvature. Thus, we conclude
that in General Relativity the effects of noncommutativity can be taken into
account by keeping the standard form of the Einstein tensor in the l.h.s.
of the field equations and introducing a modified energy-momentum tensor as a
 source in the r.h.s.
This is exactly the  gravitational analogue of the NC modification of quantum 
field theory \cite{ae0}, \cite{ae2}. \\  
It has been shown \cite{ae0}, \cite{ae2} that noncommutativity eliminates 
point-like structures in favor of smeared objects in flat spacetime.
The effect of smearing is  mathematically implemented as a ``substitution rule'' : 
position Dirac-delta function is replaced everywhere  with a Gaussian distribution 
of minimal width $\sqrt{\theta}$. Inspired by this result, we choose the mass 
density of a static, spherically symmetric, smeared, particle-like gravitational 
source as 
 
\begin{equation}
\rho_\theta\left(\,r\,\right)=
\frac{M}{\left(\,4\pi\theta\,\right)^{3/2}}\,
\exp\left(-r^2/4\theta\,\right) 
 \label{t00}
 \end{equation}

The particle mass $M$, instead of being perfectly localized at the point
is, instead, \textit{diffused} throughou a region of linear size 
$\sqrt{\theta}$. This is due to the intrinsic uncertainty encoded in
the coordinate commutator (\ref{ncx}).\\
Phenomenological results, so far, imply that
noncommutativity is not visible at presently accessible energies, i.e.
$\sqrt\theta< 10^{-16}\, cm $. Thus, at ``large distances'' one expects
minimal deviations from standard vacuum  Schwarzschild geometry. On the other 
hand, new physics is expected at distance $r\simeq \sqrt\theta$  where a non 
negligible density of energy and momentum is present. \\ 
Guided by the above considerations, we are going to look for
a static, spherically symmetric,  asymptotically Schwarzschild
solution of the Einstein equations, with (\ref{t00}) describing the energy density
of the system.  
In order to completely define the energy-momentum tensor, we rely on the
covariant conservation condition $T^{\mu\nu}\,;\,\nu=0$. 
For spherically symmetric metric this equation turns out to be 

\begin{equation}
\partial_r\, T^r{}_r= -\frac{1}{2} g^{00}\partial_r g_{00}\left(\, T^r{}_r -
T^0{}_0\,\right) - g^{\theta\theta}\partial_r g_{\theta\theta}
\left(\, T^r{}_r -T^\theta{}_\theta\,\right)
\end{equation}

To preserve  Schwarzschild-like property:
$g_{00}=-g_{rr}^{-1}$, we require 
$T^r{}_r=T^0{}_0=-\rho_\theta\left(\,r\,\right)$.
With this requirement the divergence free equation allows a solution for
$T^\theta{}_\theta$ which reads 

\begin{equation}
T^\theta{}_\theta=
- \rho_\theta\left(\,r\,\right)- \frac{r}{2}\,\partial_r\rho_\theta\left(\,r\,\right)
\end{equation}

Rather than a massive, structureless point, a source
turns out to a \textit{self-gravitating}, \textit{droplet } of 
\textit{anisotropic fluid} of density $\rho_\theta$, radial pressure 
$p_r= -\rho_\theta$ and \textit{tangential pressure} 

\begin{equation}
p_\perp = -\rho_\theta -\frac{r}{2}\,\partial_r\rho_\theta\left(\,r\,\right)\label{perp}
\end{equation}

On physical grounds, a non-vanishing radial pressure is needed to balance the inward 
gravitational pull, preventing droplet to collapse into a matter point. This is 
the basic physical effect on matter caused by spacetime noncommutativity and the 
origin of all new physics at distance scale of order  $\sqrt\theta$. By solving
Einstein equations with (\ref{t00}) as a matter source. we find the line
element \footnote[1]{We use convenient units $G_N=1$, $c=1$.}:

\begin{equation}
 ds^2 =\left(\, 1- \frac{4M}{r\sqrt{\pi}}\, \gamma(3/2 \ ,
r^2/4\theta\,)\, \right)\, 
dt^2 - \left(\, 1-\frac{4M}{r\sqrt{\pi}}\, \gamma(\, 3/2\ , r^2/4 \theta\, )\,
\right)^{-1}\, dr^2 
- r^2\,\left(\, d\vartheta^2 +\sin^2\vartheta\, d\phi^2\,\right)
\label{ncs}
\end{equation}

where $\gamma\left(3/2 \ , r^2/4\theta\, \right)$ is the lower
incomplete Gamma function:

\begin{equation}
\gamma\left(3/2\ , r^2/4\theta\, \right)\equiv
\int_0^{r^2/4\theta} dt\, t^{1/2} e^{-t}
\end{equation}

\begin{figure}[ht]
\begin{center}
\includegraphics[width=12cm,angle=0]{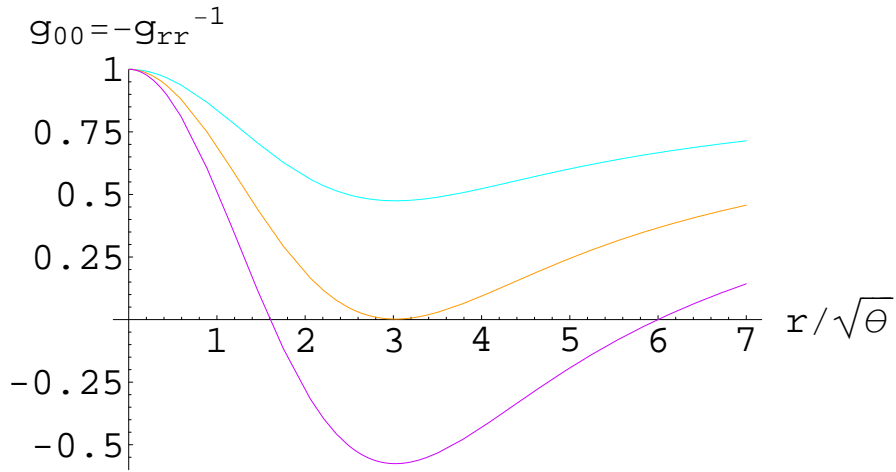}
\caption{\label{hor} $g_{rr}^{-1}$ vs $r$, for various values of $M/\sqrt\theta$.
Intercepts on the horizontal axis give radii of the event horizons.
$M= \sqrt\theta $, 
( cyan curve ) no horizon;  $M=1.9\, \sqrt\theta $, (yellow curve) one 
\textit{degenerate} horizon $r_0\approx 3.0\, \sqrt\theta$, \textit{extremal}
 black hole;  $M= 3\, \sqrt\theta $ (magenta curve) two horizons.}
\end{center}
\end{figure}

The classical Schwarzschild metric 
is obtained from (\ref{ncs})  in the limit $r/\sqrt{\theta}\to\infty $.
Eq.(\ref{t00}) leads to the mass distribution $m\left(\, r\,\right)= 
2M \,\gamma\left(3/2\ , r^2/4\theta\, \right)/\sqrt\pi $ where $M$ is
the total mass of the source.  
We recall the following properties of matter density $\rho_\theta$:

\begin{enumerate}
\item near the origin, i.e. $r<<\sqrt\theta$,
$\frac{d\rho_\theta}{dr}\simeq0\longrightarrow \rho_\theta\simeq 
\rho_\theta\left(\, 0\,\right)$   ;
\item few standard  deviations from the origin, say $r\ge 4\sqrt\theta $,
$\frac{d\rho_\theta}{dr}\simeq 0$ $\longrightarrow$ 
$\rho_\theta~\simeq~\mathrm{const. }~<<
~\rho_\theta\left(\, 0\,\right)$
\item asymptotically far away, i.e. $r>> 2M $, $\rho_\theta= 0$ .
\end{enumerate}

\begin{figure}[ht]
\begin{center}
 \includegraphics[width=12cm, angle=0 ]{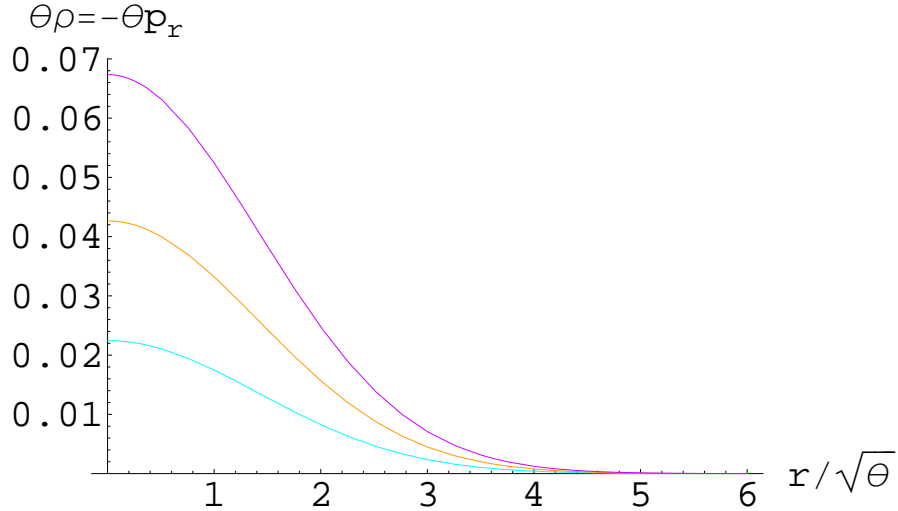}
\end{center}
\caption{\label{rho} Plot of $\theta \,\rho$ vs $r/\sqrt\theta$ for different
values of $M/\sqrt\theta$:  $M/\sqrt\theta=1$ cyan curve; $M/\sqrt\theta=1.9$
yellow curve;  $M/\sqrt\theta=3$ magenta curve. Matter distribution is
\textit{flat} both near the origin and for $r\ge 4\sqrt\theta$. }
\end{figure}

\begin{figure}[ht]
\begin{center}
 \includegraphics[width=12cm, angle=0 ]{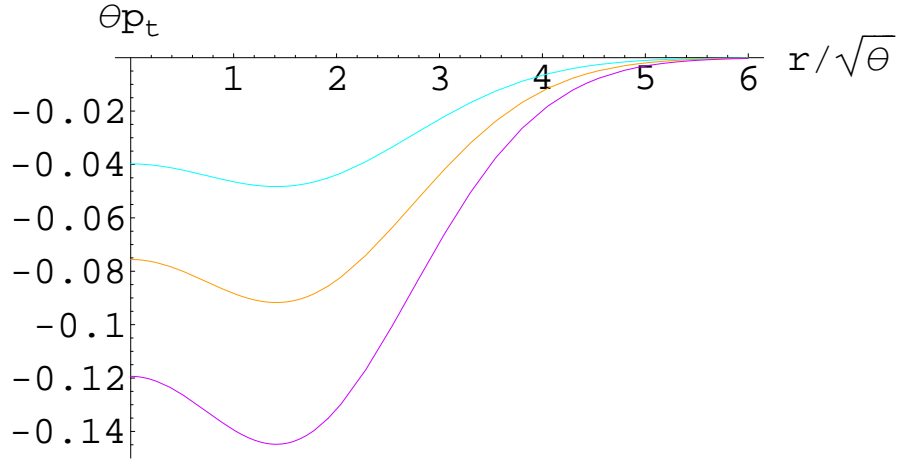}
\end{center}
\caption{\label{p} Plot of the tangential pressure
$p \theta $ vs $r/\sqrt\theta$ for different
values of $M/\sqrt\theta$.    $M/\sqrt\theta=1$ cyan curve; $M/\sqrt\theta=1.9$
yellow curve;  $M/\sqrt\theta=3$ magenta curve. The pressure is always negative,
and it is flat both near the origin and for  $r\ge 5\sqrt\theta$. The scale
has been chosen to magnify the behavior in the range $2\sqrt\theta\le r\le 
5\sqrt\theta $.}
\end{figure}

The function $p_\perp(r)$ defined in (\ref{perp}) is plotted in 
\textbf{Fig.(\ref{p})}

As far as the sign of the pressure is concerned, it is worth to
remark that we are \textit{not} dealing with the inward pressure of outer layers
of matter on the core of a ``star'', but with a  totally different  kind of
``quantum'' pressure.  It is the outward push, which is conventionally defined 
to be negative,  induced by noncommuting coordinate quantum fluctuations.
 In a simplified  picture,  quantum pressure is akin to the  cosmological 
 constant  in DeSitter universe. As a consistency check of this interpretation
 we are going to show that line element (\ref{ncs}) is well described near the
 origin by a DeSitter geometry.\\ 
The line element (\ref{ncs})  describes the geometry of a
noncommutative black hole  and should give us useful
insights about  possible noncommutative effects on Hawking radiation.\\ 
Eventual event horizon(s) can be found where  $g_{00}\left(\, r_H\,\right)=0$,
that is 

\begin{equation}
r_H= \frac{4M}{\sqrt{\pi}}\,\gamma\left(3/2\ , r^2_H/4\theta\,
\right) \label{horizon}
\end{equation}

Equation (\ref{horizon})  cannot be solved in closed form. However,
by plotting $g_{00}$ one can read  intersections with the $r$-axis and determine
 numerically the existence of horizon(s) and their radius.
Figure (\ref{hor}) shows that noncommutativity introduces new behavior
with respect to standard Schwarzschild black hole. Instead of a single event
horizon, there are different possibilities:\\
 \begin{enumerate}
 \item two distinct horizons  for $M> M_0$ (yellow curve) ;
 \item one degenerate horizon  in  $r_0=3.0\times \sqrt\theta $, with
 $M= M_0=1.9\times (\sqrt\theta) /G $
 corresponding to \textit{ extremal black hole} (cyan curve);
\item no horizon  for  $M< M_0$ (violet curve)  .
\end{enumerate}
Notice that the outer horizon has always a radius larger than the extremal
limit, i.e. $r_H\ge 3\sqrt\theta$, thus our previous approximation 
$r_H\ge 4\sqrt\theta$ is quite acceptable to get a qualitative picture of
black hole evaporation. A more refined treatment is only required close to
the extremal limit if one need precise quantitative estimates.  \\
In view of these results, there can be no black hole if the original mass is
less than the \textit{minimal mass} $M_0$. Furthermore,  contrary to the usual
case, there can be \textit{two horizons} for large masses. By increasing $M$, i.e.
for $M>> M_0$, the  \textit{inner  horizon} shrinks to zero, while the 
\textit{outer} one approaches the Schwarzschild value $r_H=2M$.\\

Equation (\ref{horizon}) can be conveniently rewritten in
terms of the upper incomplete Gamma function as

\begin{equation}
 r_H= 2M\,\left[\, 1 -\frac{2}{\sqrt{\pi}}\, \Gamma\left(\, 3/2\ ,
M^2/\theta\, \right)  \,\right] \label{horizon2}
\end{equation}

The first term in (\ref{horizon2}) is the Schwarzschild radius,
while the second term brings in $\theta$-corrections. \\
In the ``large radius'' regime  $r^2_H/4\theta>>1$  equation (\ref{horizon2}) can
be solved by iteration. At  the first order  in $M/\sqrt\theta$, we find 

\begin{equation}
r_H = 2M\,\left(\, 1 - \frac{M}{\sqrt{\pi\theta}}\,
  e^{-M^2/\theta}\,\right)\label{2m}
\end{equation}

The effect of noncommutativity is exponentially small, which is
reasonable to expect since at large distances spacetime can be
considered as a smooth classical manifold. 
 On the other hand, at short distance one expects significant changes due to the
spacetime fuzziness, i.e. the ``quantum geometry''  becomes
important $r_H\simeq\sqrt\theta  $. \\
Let us now consider the black hole temperature:

\begin{eqnarray}
T_H&&\equiv 
\left(\,\frac{1}{4\pi}\, \frac{d g_{00}}{dr}\right)_{r=r_H}
\nonumber\\
&&=
\frac{1}{4\pi\,r_H}\left[\, 1 
 -\frac{r^3_H}{4\,\theta^{3/2}}\,
  \frac{e^{-r^2_H/4\theta}}{\gamma\left(\, 3/2\ ; r^2_H/4\theta \right)} 
\,\right]
\label{thnc}
\end{eqnarray}

where, we expressed $M$ in terms of $r_H$ from the horizon equation
(\ref{horizon}).
For large black holes, i.e.  $r_H^2/4\theta>>1$,   and  one recovers the
standard result for the Hawking temperature

\begin{equation}
T_H= \frac{1}{4\pi\, r_H}  \label{th}
\end{equation}

\begin{figure}[ht]
\begin{center}
 \includegraphics[width=12cm, angle=0 ]{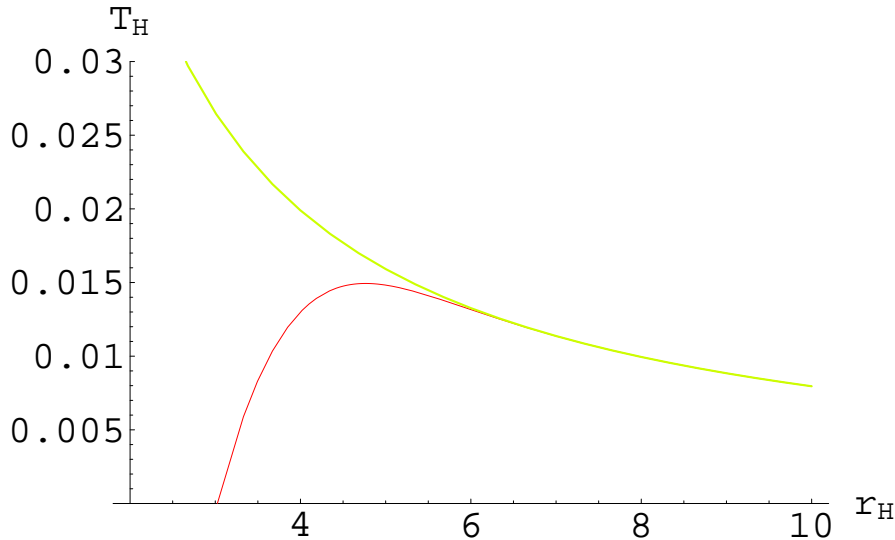}
\end{center}
\caption{\label{T} Plot of $T_H$ vs $r_H$, in $\sqrt\theta$ units. yellow curve
is the plot of (\ref{thnc}):
$T_H=0$ for $r_H=r_0=3.0\sqrt{\theta}$,i.e. for the extremal black hole,
 while the maximum temperature $T_H\simeq 0.015\times
1/\sqrt{\theta}$ corresponds to a mass $M\simeq 2.4 \times\sqrt{\theta}$. 
For comparison, we plotted in yellow the standard Hawking temperature.  The two
temperatures coincide for $r_H > 6\,\sqrt\theta$.}
\end{figure}

At the initial state of radiation the black hole temperature increases while
the horizon radius is decreasing. It is crucial  to investigate what
happens as $r_H\to \sqrt\theta$. In the standard (~commutative~) case 
  $T_H$ diverges and this puts limit on the
validity of the conventional  description of Hawking radiation.
Against this scenario,  temperature (\ref{thnc}) includes noncommutative effects
which are relevant at distances comparable to  $\sqrt\theta$. 
Behavior of the temperature $T_H$ as a function of the horizon radius is
plotted in \textbf{Fig.(\ref{T})}.

In the region $r_H\simeq \sqrt\theta $, $T_H$ deviates from the standard hyperbola
(\ref{th}). Instead of exploding with shrinking $r_H$,
$T_H$ reaches a maximum\footnote[2]{We are in the
range of distances, $r_H\ge 4\sqrt\theta $, where our approximation is physically
reliable.}  in $r_H\simeq 4.7\sqrt\theta $ 
corresponding to a
mass $M\approx 2.4\, \sqrt\theta/ G_N$, then quickly drops to zero for 
$r_H=r_0=3.0\sqrt{\theta}$ corresponding to the radius of the extremal
black hole in figure (\ref{hor}). In the region $r< r_0 $ there is no black
hole  and the corresponding temperature cannot be defined.
As a summary of the results,
the emerging  picture of non commutative black hole is that 
for $M >> M_0$ the temperature is given by the Hawking temperature
(\ref{th}) with negligibly small exponential corrections, 
and  increases, as the mass  
is radiated away. $T_H$ reaches a maximum value at $M= 2.4\, \sqrt\theta$ and 
then drops  as $M$ approaches $M_0$. When $M=M_0$, $T_H=0$, event horizon is 
degenerate, and we are left with a ``frozen'' extremal black hole. \\
The temperature behavior shows that  noncommutativity plays the same role in
General Relativity as in Quantum Field Theory, i.e.  removes short distance
divergences.\\
At this point, important issue of Hawking radiation back-reaction should be 
discussed. In commutative case one expects relevant back-reaction effects 
during the terminal stage of evaporation  because of huge increase of
temperature \cite{backr}, \cite{swave}. As it has been shown, the role of 
noncommutativity is to cool down the black hole in the final stage.
As a consequence, there is a suppression of quantum back-reaction since 
the black hole emits less and less  energy. Eventually, 
back-reaction may be important during the maximum temperature phase. In order to 
estimate its importance in this region, let us look at the thermal energy 
$E=T_H\simeq 0.015\, /\sqrt{\theta}$ 
and the total mass $M\simeq 2.4\,\sqrt{\theta}\,M_{Pl.}^2  $. In order to
have significant back-reaction effect $ T_H^{Max}$ should be of the same order
of magnitude as $M$. This condition leads to the estimate 

\begin{equation}
\sqrt{\theta}\approx 0.2\, l_{Pl.}\sim 10^{-34}\, cm \label{stima}
\end{equation}

Expected values of $\sqrt{\theta}$ are well above the Planck length 
$l_{Pl.}$ and (\ref{stima}) indicates that  back-reaction effects are 
suppressed even if $\sqrt{\theta}\approx 10\, l_{Pl.}$ and
$T_H^{Max}\approx 10^{16}\, GeV$.  For this reason
we can safely use unmodified form of the metric (\ref{ncs}) 
during all the evaporation process.\\ 
Finally, we would like to clarify what happens if the starting object has
mass smaller than $M_0$. It has been shown that in this case there is no
event horizon. Could it be a \textit{naked singularity}? 
 We are going to show  that  this is not the case.  
In order to establish the geometrical picture for sub-minimally massive objects
we are going  to study the curvature scalar near $r=0$. In case of  naked
singularity one should obtain divergent curvature, $R\to\infty$. 
Short distance behavior of $R$ is obtained by tracing the Einstein equations.
One finds

\begin{equation}
 R\left(\, 0\,\right)=\frac{4M}{\sqrt\pi\, \theta^{3/2}} 
  \label{ricci0}
\end{equation}

For $r<<\sqrt\theta$ the curvature is actually \textit{constant and
positive}.  An eventual naked singularity is replaced 
 by a DeSitter, \textit{regular} geometry around the origin.  Earlier attempts
 to avoid curvature singularity at the origin of
 Schwarzschild metric have been made by matching DeSitter
 and Schwarzschild geometries both along time-like \cite{cosmo}, and space-like 
 matter shells \cite{frolov}, or by cleverly guessing regular
 black hole geometries and deriving the corresponding  energy momentum tensor
 from the Einstein equations \cite{regbh}. Our approach is ``dual'' to the
 one in \cite{regbh} as we start from an energy-momentum tensor originating from
 noncommutative geometry and with this source we solve Einstein equation. \\
As a conclusion, the results derived in this letter show that the coordinate 
coherent state approach to noncommutative effects can cure the singularity
problems at the terminal stage of black hole evaporation. \\
We have shown that noncommutativity, which is an intrinsic property of the manifold
itself, can be introduced in General Relativity by modifying the matter source.
Energy-momentum tensor needed for this description is of the ideal fluid form
and requires a non-trivial pressure. In spite of complicated equation of state
it can be studied in the regions of interest and new black hole behavior is 
discovered in the region $r\simeq \sqrt\theta$.
Specifically,
we have shown that there is a minimal mass $M_0= 1.9\, \sqrt{\theta}$ to
which a black hole can decay through Hawking radiation.

The reason why it does not end-up into
a naked singularity is due to the finiteness of the  curvature at the origin.
The everywhere regular  geometry and the residual  mass $M_0$ are both 
manifestations of the Gaussian de-localization of the source in the 
noncommutative spacetime.
On the thermodynamic side,  the same kind of regularization  takes place 
eliminating the divergent behavior of Hawking temperature. 
As a consequence there is a maximum temperature
that the black hole can reach before cooling down to absolute zero. 
As already anticipated in the introduction,  noncommutativity regularizes 
divergent quantities in the final stage of
black hole evaporation in the same way it cured UV infinities
in noncommutative quantum field theory.\\
We have also estimated that back-reaction does not modify the original metric 
in a significant manner.

\acknowledgements
We would like to thank Dr. Thomas G. Rizzo for a careful reading of the paper
and for pointing out a couple of misprints.\\
One of us, P.N., thanks the ``Dipartimento di Fisica Teorica, Universit\`a di 
Trieste'' and PRIN-COFIN project 2004 ``Metodi matematici per le teorie cinetiche''
for financial support.


\begin{thebibliography}{99}
\bibitem{hawking} S. W. Hawking,\ Comm.\ Math.\ Phys.\ \textbf{43}\
199\ (1975)
\bibitem{paddy} T. Padmanabhan,\ Phys.\ Rep.\ \textbf{406}\ 49\ (2005)
\bibitem{corr} L. Susskind,\ Phys.\ Rev.\ Lett.\ \textbf{71}\ 2367\ (1993)
\bibitem{sw} E. Witten, Nucl.\ Phys.\ B \textbf{460}\ 335 \ (1996);
        N. Seiberg,  E. Witten,\  JHEP\ \textbf{9909}\   032\  (1999)
\bibitem{sny}H. S. Snyder, \ Phys.\ Rev.\ \textbf{71}\ 38\ (1947)
\bibitem{ae0} A. Smailagic, E. Spallucci,\  J.\ Phys.\ \textbf{A36}\  L467\ (2003)
\bibitem{ae2} A. Smailagic, E. Spallucci,\  J.\ Phys.\
\textbf{A36}\  L517\ (2003)
\bibitem{casino} M. Chaichian, A. Demichev, P. Presnajder, \
   Nucl.\ Phys.\ B\textbf{567}\ (2000)\ 360;\ 
   S. Cho, R. Hinterding, J. Madore, H. Steinacker,\ 
	Int.\ J.\  Mod.\ Phys.\ D\textbf{9}\ (2000)\ 161
\bibitem{ae} A. Smailagic, E. Spallucci, \ 
	 J.\ Phys.\  \textbf{A37}\ 7169\ (2004)
\bibitem{berna} P. Nicolini,\  A. Smailagic and E. Spallucci,\ 
\textit{ ``The fate of radiating black holes in noncommutative geometry}
 Proceedings of EPS 13 General Conference  "Beyond Einstein,
  Physics for the 21st Century" July 11-15, 2005, University of Bern, Bern, 
  Switzerland; e-Print Archive: hep-th/0507226
\bibitem{ag} A. Gruppuso, 
	 J.\ Phys.\ \textbf{A38}\ 2039\ (2005)
\bibitem{piero2005}	 
P. Nicolini\  J.\ Phys.\ \textbf{A38}\ L631\ (2005)
\bibitem{noi4} A.\ Smailagic, E.\ Spallucci,\ Phys.\ Rev.\ \textbf{D65} (2002)
107701\\
A.\ Smailagic, E.\ Spallucci,\ J.\ Phys.\ \textbf{A35}\ L363\ (2002)
\bibitem{backr}	 R. Balbinot, A. Barletta, \ Class.\ Quant.\ Grav.
\textbf{6}\ 195,\ 1989;
\bibitem{swave} 
	R. Balbinot, A. Fabbri, V. Frolov,  P. Nicolini, P. J. Sutton,
	A. Zelnikov,\  Phys.\ Rev.\ \textbf{D63}\  084029\  (2001);\\
R. Balbinot, A. Fabbri, P. Nicolini, P. J. Sutton,\ 
	 Phys.\ Rev.\ \textbf{D66}\  024014\  (2002)
\bibitem{cosmo} 
A. Aurilia, G. Denardo, F. Legovini, E. Spallucci
Phys.\ Lett.\ \textbf{B147}\ 258 \ (1984)
A. Aurilia, G. Denardo, F. Legovini, E. Spallucci
Nucl.\ Phys.\ B \textbf{252}\ 523\ (1985)\\
A. Aurilia , R.S. Kissack, R. Mann\ 
Phys.\ Rev.\ \textbf{D35}\  2961\  (1987)
\bibitem{frolov} V. P. Frolov, M.A. Markov, V.F. Mukhanov
     Phys.\ Rev.\ \textbf{D41}\ 383\ (1990)
\bibitem{regbh}	
D. A. Easson \ JHEP \textbf{0302}\ 037\ (2003) \\
S. A. Hayward
``\textit{Formation and evaporation of regular black holes}''
e-Print Archive: gr-qc/0506126
	\end{thebibliography}
	\end{document}